\documentclass[11pt]{article}

\usepackage[preprint]{acl}

\usepackage{times}
\usepackage{latexsym}
\usepackage{amsmath}
\usepackage{amsthm} 
\usepackage{amssymb}

\usepackage{amsfonts}  
\usepackage{algorithm}
\usepackage{algpseudocode} 

\usepackage{enumitem}

\usepackage{xcolor}
\usepackage{tcolorbox}
\usepackage{booktabs}

\definecolor{evoblue}{RGB}{235, 245, 255}
\definecolor{evodark}{RGB}{0, 50, 120}
\definecolor{initgray}{RGB}{245, 245, 245}
\definecolor{initred}{RGB}{150, 0, 0}

\usepackage[T1]{fontenc}

\usepackage[utf8]{inputenc}

\usepackage{microtype}

\usepackage{inconsolata}

\usepackage{graphicx}
\usepackage{multirow}
\usepackage{pifont}

\theoremstyle{definition}

\title{MulVul: Retrieval-augmented Multi-Agent Code Vulnerability Detection \\ via Cross-Model Prompt Evolution}

\author{
	Zihan Wu,
	Jie Xu,
	Yun Peng,
	Chun Yong Chong \and
	Xiaohua Jia
}

\begin{document}
\maketitle

\begin{abstract}
	Large Language Models (LLMs) struggle to automate real-world vulnerability detection due to two key limitations: the heterogeneity of vulnerability patterns undermines the effectiveness of a single unified model, and manual prompt engineering for massive weakness categories is unscalable.  
	To address these challenges, we propose \textbf{MulVul}, a retrieval-augmented multi-agent framework designed for precise and broad-coverage vulnerability detection. MulVul adopts a coarse-to-fine strategy: a \emph{Router} agent first predicts the top-$k$ coarse categories and then forwards the input to specialized \emph{Detector} agents, which identify the exact vulnerability types. Both agents are equipped with retrieval tools to actively source evidence from vulnerability knowledge bases to mitigate hallucinations. 
	Crucially, to automate the generation of specialized prompts, we design \emph{Cross-Model Prompt Evolution}, a prompt optimization mechanism where a generator LLM iteratively refines candidate prompts while a distinct executor LLM validates their effectiveness. This decoupling mitigates the self-correction bias inherent in single-model optimization. 
	Evaluated on 130 CWE types, MulVul achieves 34.79\% Macro-F1, outperforming the best baseline by 41.5\%. Ablation studies validate cross-model prompt evolution, which boosts performance by 51.6\% over manual prompts by effectively handling diverse vulnerability patterns.
\end{abstract}

\section{Introduction}  
Code vulnerabilities pose a fundamental threat to software reliability and security, leading to software crashes and service interruptions~\cite{peng2024domain}. As modern software systems grow in complexity, manual code auditing has become increasingly expensive, time-consuming, and error-prone, motivating the need for automated vulnerability detection~\cite{ghaffarian2017software}.

Recent advances in large language models (LLMs) have sparked interest in their application to vulnerability detection~\cite{Peng2025iCodeReviewer,zhou2025large}. Previous efforts primarily focused on single-model approaches, where a unified model is fine-tuned or prompted to identify all vulnerability types simultaneously~\cite{gao2025texttt,lin2025large}. However, vulnerability patterns are highly heterogeneous~\cite{chakraborty2021deep}. For example, buffer overflows require reasoning about pointer arithmetic and memory bounds, while injection attacks require tracking how untrusted inputs flow into sensitive operations. As a result, a single unified detector struggles to capture these diverse, type-specific patterns within a shared latent space, leading to missed vulnerabilities or high false alarm rates. 

\begin{figure}[t]
	\centering
	\includegraphics[width=\linewidth]{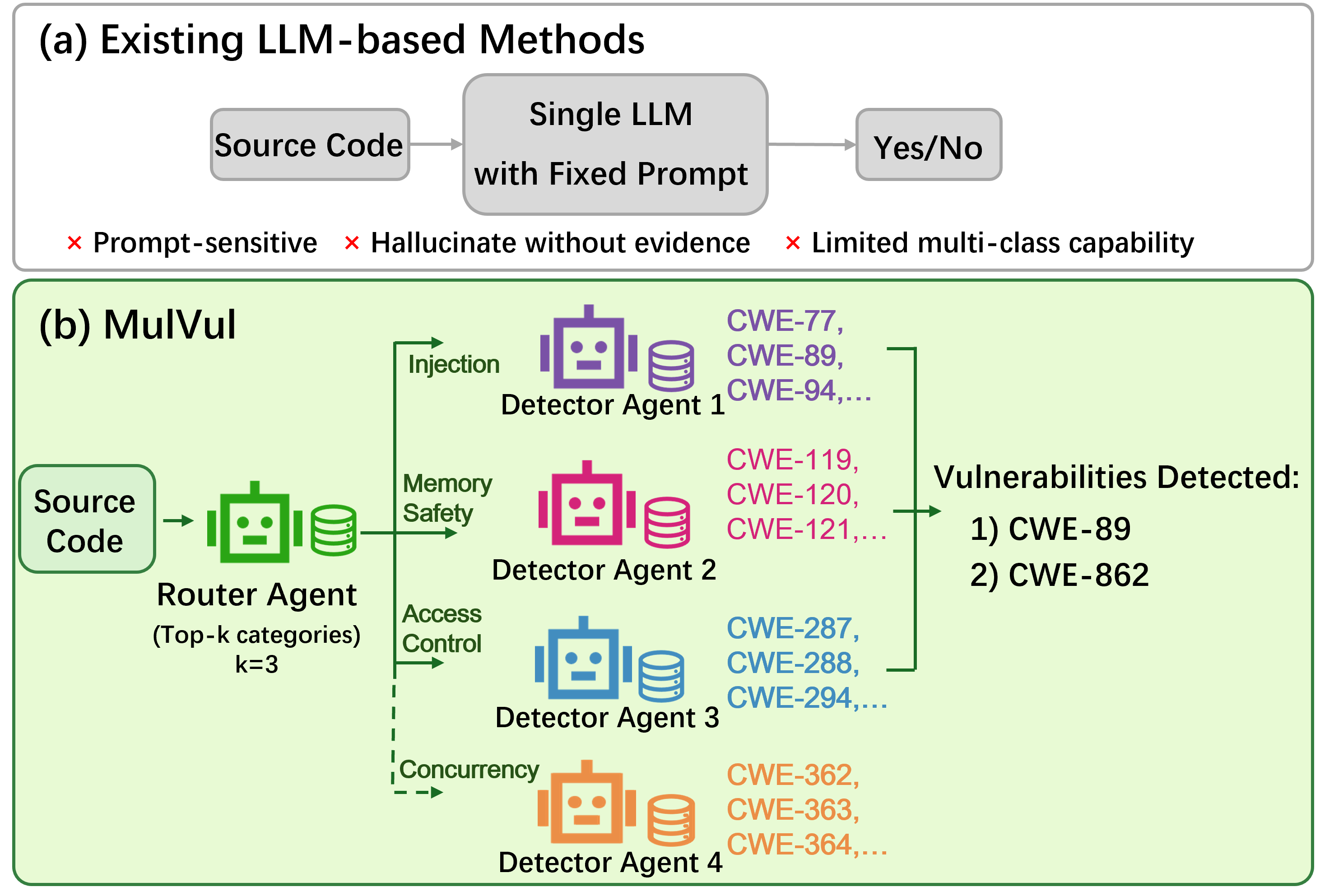}
	\caption{Comparison between MulVul and existing LLM-based vulnerability detection methods. (a) Existing methods rely on fixed prompts and lack external grounding. (b) MulVul adopts a coarse-to-fine, retrieval-augmented multi-agent framework for multi-type vulnerability detection.} 
	\label{fig:cmp}
\end{figure} 

Inspired by the success of multi-agent systems that decompose complex tasks into specialized components~\cite{wu2024autogen}, a question arises: \emph{Can a multi-agent architecture enhance multi-class vulnerability detection by routing inputs to specialized experts?} 

It is challenging to apply a multi-agent architecture for broad-coverage vulnerability detection.
First, it is computationally prohibitive to invoke a specialized agent for every vulnerability type. Real-world systems involve hundreds of Common Weakness Enumeration (CWE) entries~\cite{cwe}. To ensure comprehensive coverage, querying every corresponding agent for each input creates an impractical inference burden.
Second, manual prompt engineering becomes unscalable in multi-agent architectures. Unlike unified models, each specialized agent requires a unique instruction to capture distinct, fine-grained patterns of vulnerability. Manually optimizing prompts for such a vast number of agents is not feasible.
Third, multi-agent LLM systems can amplify hallucinations. Evidence of vulnerabilities is often dispersed across complex control flows, causing agents to reason under uncertainty. If an individual agent hallucinates a flaw, this error can cascade through inter-agent communication, distorting the final consensus~\cite{hong2023metagpt}. 

To address these challenges, we propose \textbf{MulVul}, a retrieval-augmented multi-agent framework equipped with cross-model prompt evolution for vulnerability detection. Figure~\ref{fig:cmp} contrasts prior methods with MulVul.  
MulVul adopts a coarse-to-fine Router-Detector architecture aligned with the hierarchical structure of CWE~\cite{cwe}. A \emph{Router agent} first predicts the Top-$k$ coarse categories, and only the corresponding category-specific \emph{Detector agents} are invoked to identify fine-grained vulnerability types in that category. This selective activation drastically reduces inference costs while maintaining high recall.
Crucially, to solve the scalability bottleneck of prompt engineering, MulVul employs a Cross-Model Prompt Evolution mechanism for prompt optimization. A generator LLM (e.g., Claude) iteratively proposes prompt candidates, while an executor LLM (e.g., GPT-4o) evaluates their fitness. By decoupling prompt generation from evaluation across different LLMs, MulVul mitigates the self-correction bias inherent in single-model optimization, yielding robust and highly specialized prompts.
To further mitigate hallucinations, agents actively query evidence from a SCALE-structured vulnerability knowledge base~\cite{wen2024scale} to ground their reasoning. Detectors operate in isolation to prevent error amplification across agents.

Experiments on the PrimeVul benchmark establish MulVul as the new state-of-the-art. Evaluated across 130 CWE types, MulVul achieves a Macro-F1 of 34.79\%, surpassing the best baseline by 41.5\%. With cross-model prompt evolution, MulVul significantly reduces false positives, ensuring detection accuracy.

The contributions are summarized as follows:
\begin{itemize}  
	\item We propose MulVul, a novel retrieval-augmented multi-agent framework for multi-class vulnerability detection. By enabling specialized agents with tool-augmented reasoning, MulVul effectively handles vulnerability heterogeneity while balancing computational efficiency with detection coverage.
	\item We design a Cross-Model Prompt Evolution mechanism that automatically optimizes the prompts of specialized agents. By separating generation from execution, this approach mitigates self-correction bias and solves the scalability challenge of manual prompt engineering.
	\item  Comprehensive experiments show that MulVul significantly outperforms baselines with a 34.79\% Macro-F1. Ablation studies confirm that our evolutionary mechanism boosts performance by 51.6\% over manual prompts, demonstrating its critical role in handling diverse vulnerability patterns.
\end{itemize}

\section{Related Work}\label{sec:related}
\noindent\textbf{Learning-based vulnerability detection.}
Learning-based vulnerability detection has progressed from early deep learning frameworks (e.g., VulDeePecker~\cite{li2018vuldeepecker}) to neural network models that learn code representations with sequence and graph encoders~\cite{zhou2019devign, li2021sysevr, chakraborty2021deep}, and more recently to pre-trained code models such as GraphCodeBERT~\cite{guo2021graphcodebert} and UniXcoder~\cite{guo2022unixcoder}. Recently, LLMs have dominated the field due to their strong code understanding capabilities~\cite{zhou2025large}. However, existing single-model approaches face a critical challenge: a unified detector often struggles to simultaneously capture the diverse and fine-grained patterns of varying vulnerability types~\cite{lin2025large, sheng2025llms}. While general-purpose multi-agent frameworks (e.g., AutoGen~\cite{wu2024autogen}, MetaGPT~\cite{hong2023metagpt}) show promise in task decomposition,  they have not been tailored to multi-class vulnerability detection under tight cost and reliability constraints. MulVul addresses this challenge by proposing a coarse-to-fine strategy that first performs coarse-grained routing and then type-specialized identification.

\noindent\textbf{Prompt engineering and optimization for LLMs.}
To reduce the reliance on manual prompt engineering~\cite{wei2022chain}, automatic optimization strategies have emerged, treating prompt generation as a search or optimization problem, such as APE~\cite{zhou2022large}, EvoPrompt~\cite{guo2024evoprompt}, and OPRO~\cite{yang2023large}. A major limitation of these methods is their reliance on a single backbone for both generation and evaluation, which risks overfitting to model-specific biases and limits transferability across LLMs. We address this by proposing Cross-Model Prompt Evolution, which decouples the generator and executor. This separation provides unbiased feedback, facilitating the discovery of robust instructions that generalize more effectively across vulnerability types.

\noindent\textbf{Retrieval-augmented generation and hallucination mitigation.}
Retrieval-augmented generation (RAG) effectively grounds LLMs to mitigate hallucinations~\cite{lewis2020rag}, with applications extending to code completion and repair~\cite{lu2022reacc}. However, standard code retrieval often focuses on syntactic similarity, which is insufficient for distinguishing subtle security flaws. MulVul advances this by leveraging SCALE-based structured semantic representations~\cite{wen2024scale} and implementing a contrastive retrieval strategy. The Router utilizes broad evidence to identify categories, while Detectors utilize contrastive example retrieval to distinguish vulnerabilities.

\section{Preliminaries and Problem Definition}\label{sec:preliminary}
\subsection{Common Weakness Enumeration (CWE)} \label{sec:cwe}
The CWE taxonomy~\cite{cwe} organizes software vulnerabilities hierarchically. We focus on a two-level structure comprising $M$ coarse-grained categories $\mathcal{C} = \{c_1, \dots, c_M\}$ (e.g., Memory Buffer Errors, Injection). Each category $c_m$ contains a set of fine-grained vulnerability types $\mathcal{Y}_m$ (e.g., CWE-119 Buffer Overflow, CWE-125 Out-of-bounds Read under Memory Buffer Errors). 

We define the complete label space as $\mathcal{Y} = \{y_0, y_1, \dots, y_K\}$, where $y_0$ denotes non-vulnerable code and $\{y_1, \dots, y_K\} = \bigcup_{m=1}^{M} \mathcal{Y}_m$, where $\mathcal{Y}_1, \dots, \mathcal{Y}_M$ are pairwise disjoint. 

\subsection{LLM-based Code Vulnerability Detection}
Given an LLM $\mathcal{M}$ with frozen parameters, we formulate vulnerability detection as a retrieval-augmented generation task. The input consists of a code snippet $x \in \mathcal{X}$ and a textual prompt $p$. Since real-world code may contain multiple vulnerabilities, we adopt a multi-class formulation where the system outputs a prediction set $\hat{\mathcal{Y}} \subseteq \mathcal{Y}$. In practice, the LLM generates structured outputs (e.g., a list of predicted CWE types), which are parsed to obtain $\hat{\mathcal{Y}}$. As $\mathcal{M}$ remains frozen, detection performance relies heavily on the prompt $p$, which serves as the optimizable variable.

\subsection{SCALE: Structured Code Representation}\label{sec:prelim_scale}
To capture code semantics and execution flow, SCALE~\cite{wen2024scale} constructs a Structured Comment Tree for vulnerability detection. Given source code $x$, SCALE uses LLMs to generate natural-language comments attached to AST nodes, then applies structured rules to encode control-flow sequences, yielding $T(x)=\mathrm{SCALE}(x)$.

\subsection{Problem Formulation}\label{sec:problem_formulation}
Given a code snippet $x \in \mathcal{X}$, our goal is to design a multi-agent system $\mathcal{A}$ that outputs $\hat{\mathcal{Y}} = \mathcal{A}(x) \subseteq \mathcal{Y}$. The system should achieve: (i) high-precision detection, (ii) robustness across LLM backbones, and (iii) computational efficiency.

\begin{figure*}[t]
	\centering
	\includegraphics[width=0.99\linewidth]{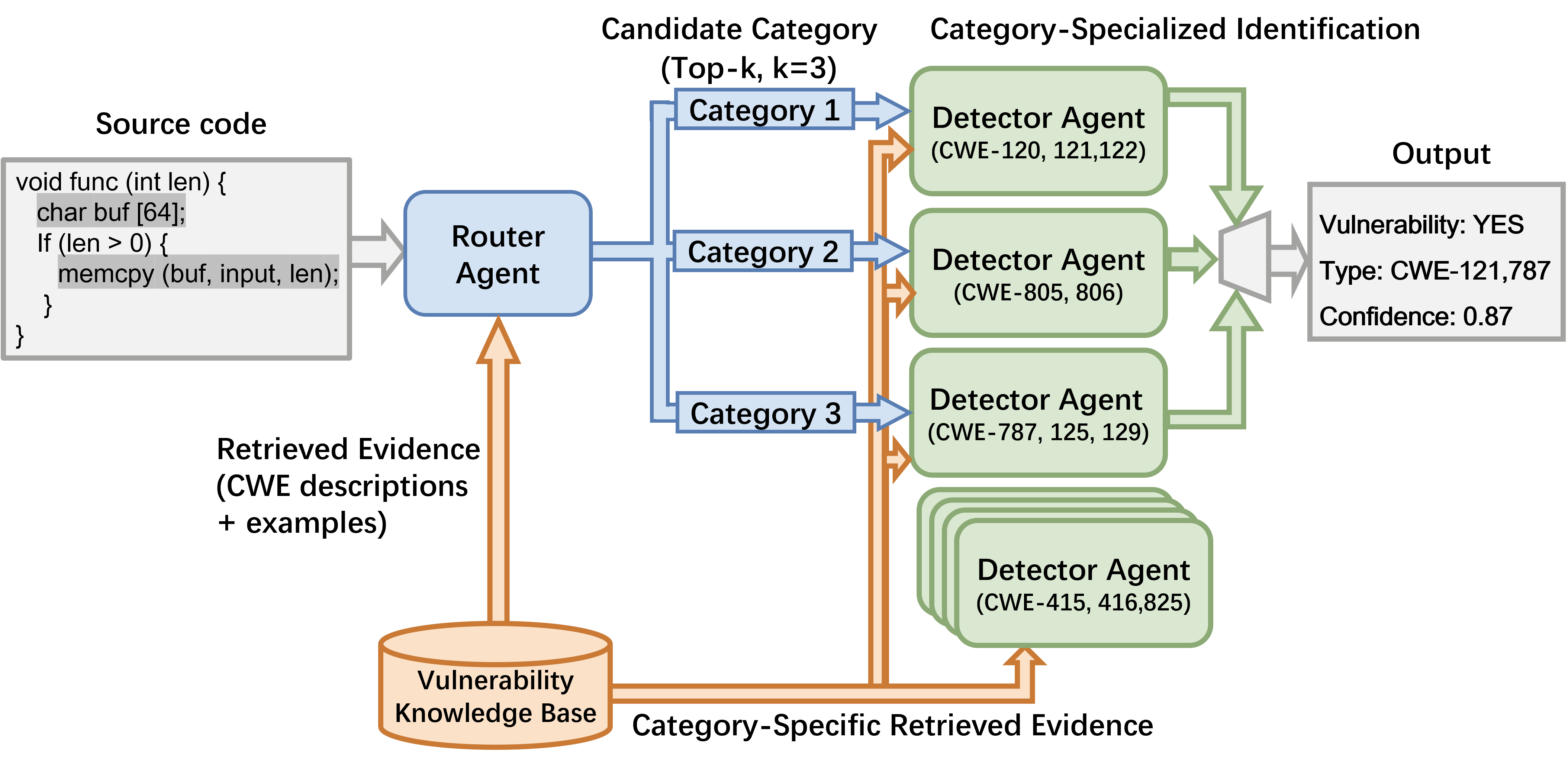}
	\caption{Overview of MulVul for vulnerability detection. The router agent first selects top-$k$ candidate vulnerability categories, and category-specific detector agents then perform fine-grained identification with retrieved CWE-specific evidence.}
	\label{fig:framework}
\end{figure*}

\section{Method}\label{sec:method}
\subsection{Overview of MulVul}
MulVul operates in two phases: offline preparation and online detection.

During offline preparation, MulVul first constructs a vulnerability knowledge base $\mathcal{K}$ by converting labeled samples into SCALE representations~\cite{wen2024scale}. MulVul then employs cross-model prompt evolution to optimize prompts for Router and Detector agents. Specifically, we use two separate LLMs with distinct roles: a generator LLM $\mathcal{M}_{\text{evo}}$ (e.g., Claude) that proposes and mutates candidate prompts, and an executor LLM $\mathcal{M}_{\text{exec}}$ (e.g., GPT-4o) that runs the Router/Detector agents and returns performance feedback. Through this process, the Router agent obtains a prompt optimized for category-level recall, while each Detector agent receives a prompt tailored for precise fine-grained identification.

During online detection, MulVul adopts a coarse-to-fine Router-Detector architecture, as illustrated in Figure~\ref{fig:framework}. Given a code snippet $x$, a Router agent first actively invokes an analysis tool to retrieve evidence from $\mathcal{K}$ and predicts the top-$k$ categories. Only the corresponding Detector agents are then invoked, each employing specialized contrastive retrieval tools to identify the exact vulnerability. Each Detector operates in isolation without inter-agent communication, avoiding error amplification.

 \subsection{Offline Preparation} \label{sec:prompt_evolution}
The offline phase 1) constructs the retrieval infrastructure and 2) optimizes prompts for Router and Detector agents. 

\subsubsection{Knowledge Base Construction}
We construct a vulnerability knowledge base $\mathcal{K}$ to provide grounding evidence for both Router and Detector agents. Given the training set $\mathcal{D}_{tr} = \{(x_i, y_i)\}_{i=1}^{N}$ where $x_i$ is a code snippet and $y_i \in \mathcal{Y}$ is its vulnerability label, we convert each sample into its SCALE representation $T(x_i)$ following~\cite{wen2024scale}. We index all transformed samples to form the knowledge base: 
\begin{equation}
	\mathcal{K} = \{(T(x_i), y_i)\}_{i=1}^{N}
\end{equation}
For efficient retrieval, we embed each SCALE representation $T(x_i)$ with UniXcoder~\cite{guo2022unixcoder} and perform nearest-neighbor search by cosine similarity. We partition the knowledge base into a clean pool $\mathcal{K}_0$ (entries labeled $y_0$) and category-specific vulnerability pools $\{\mathcal{K}_m\}_{m=1}^M$ (entries whose CWE category is $c_m$, i.e., $\mathcal{K}_m = \{(T(x_i), y_i) \in \mathcal{K} \mid y_i \in \mathcal{Y}_m\}$). For Detector $m$, we denote $\mathcal{K}_{\neg m} = \bigcup_{j \neq m} \mathcal{K}_j$ as the set of out-of-category vulnerabilities. 

During detection, the Router agent invokes the global retrieval tool to access evidence across categories, while each Detector agent employs the contrastive tool to source in-category and hard-negative examples. During the training phases (Stage I and II), when retrieving evidence for a training sample $x_i \in \mathcal{D}_{tr}$, we strictly exclude $x_i$ itself to prevent data leakage.

\subsubsection{Cross-Model Prompt Evolution}
\begin{figure}[t]
	\centering
	\includegraphics[width=0.99\linewidth]{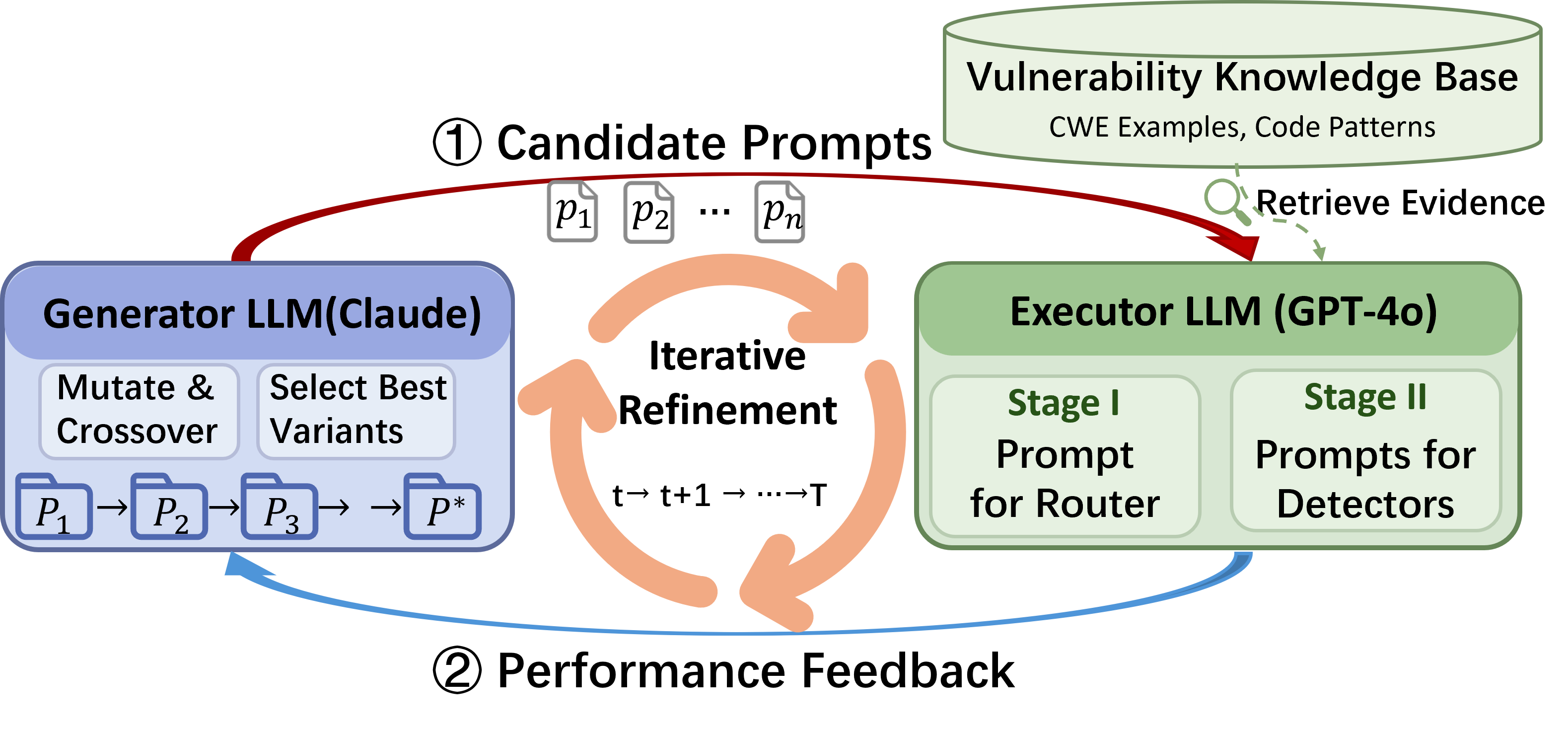}
	\caption{Illustration of the Cross-Model Prompt Evolution Process. The generator LLM $\mathcal{M}_{\text{evo}}$ (Claude ) proposes and mutates prompts, while the executor LLM $\mathcal{M}_{\text{exec}}$ (GPT-4o) evaluates their fitness.}
	\label{fig:training}
\end{figure}

As illustrated in Figure~\ref{fig:training}, the key idea is to decouple prompt generation from execution across different LLMs: an evolution model $\mathcal{M}_{\text{evo}}$ generates and refines candidate prompts, while an execution model $\mathcal{M}_{\text{exec}}$ evaluates them on the detection task. Both $\mathcal{M}_{\text{evo}}$ and $\mathcal{M}_{\text{exec}}$ remain frozen throughout the optimization process; only the textual prompts are evolved. This separation enhances exploration of the prompt space: since $\mathcal{M}_{\text{evo}}$ and $\mathcal{M}_{\text{exec}}$ have different internal biases, mutations proposed by $\mathcal{M}_{\text{evo}}$ are less likely to exploit superficial patterns, reducing premature convergence to locally optimal prompts.

Algorithm~\ref{alg:evolution} presents the optimization procedure, which proceeds in two stages. Router optimizes Recall@$k$ for coverage; Detectors optimize F1 for precision-recall balance.

\textbf{Stage I: Router Prompt Optimization.}
We initialize $n$ candidate prompts $\mathcal{P}_R$ using manually designed templates that specify the task format and output structure. In each generation, every prompt $p \in \mathcal{P}_R$ is executed by $\mathcal{M}_{\text{exec}}$ on training samples with retrieved evidence from $\mathcal{K}$. We use Recall@$k$ as the fitness function because the Router aims to ensure the correct category is included in top-$k$ predictions, avoiding early filtering of true vulnerabilities. The evolution model $\mathcal{M}_{\text{evo}}$ then evolves the prompts through the \textsc{Evolve} procedure (Algorithm~\ref{alg:evolve}): high-fitness prompts are retained, and new candidates are generated via LLM-driven mutation (e.g., rephrasing instructions, adding constraints, adjusting output format). Throughout evolution, fitness is computed on the training set $\mathcal{D}_{tr}$. After all iterations are complete, we evaluate each generation's best prompt (tracked during training) on the held-out validation set $\mathcal{D}_{val}$, and select the one with the highest Recall@$k$ as $p_R^*$.
 
\textbf{Stage II: Detector Prompt Optimization.} 
We optimize each Detector prompt independently and in parallel. For category $c_m$, we construct $\mathcal{D}_{tr}^{(m)}$ and $\mathcal{D}_{val}^{(m)}$ with in-category positives, clean negatives, and out-of-category vulnerabilities (hard negatives). Each Detector is evaluated with F1 score using evidence from $\mathcal{K}_m$ (positives), $\mathcal{K}_0$ (clean), and $\mathcal{K}_{\neg m}$ (other categories). The evolution mirrors Stage I, and parallelization across $M$ categories ensures efficiency. 
 
\begin{algorithm}[t]
	\caption{Cross-Model Prompt Evolution}
	\label{alg:evolution}
	\begin{algorithmic}[1]
		\Require $\mathcal{M}_{\text{evo}}$, $\mathcal{M}_{\text{exec}}$, $\mathcal{K}$, $\mathcal{D}_{tr}$, $\mathcal{D}_{val}$, Categories $M$, Iterations $T$
		\Ensure Optimized prompts $p_R^*$, $\{p_m^*\}_{m=1}^{M}$
		
		\Statex \textbf{// Stage I: Router Prompt Optimization}
		\State Initialize prompts $\mathcal{P}_R \leftarrow \{p_1, \dots, p_n\}$
		\For{$t = 1$ to $T$}
		\State $\mathcal{S} \leftarrow \{ \text{Recall@}k(p, \mathcal{M}_{\text{exec}}, \mathcal{D}_{tr}) \mid p \in \mathcal{P}_R \}$ 
		\State $\mathcal{P}_R \leftarrow \textsc{Evolve}(\mathcal{P}_R, \mathcal{S}, \mathcal{M}_{\text{evo}})$
		\State Track best prompt $p_{\text{best}}^{(t)}$ based on $\mathcal{S}$
		\EndFor
		\State Let $\mathcal{P}_{\text{best}} = \{p_{\text{best}}^{(1)}, \dots, p_{\text{best}}^{(T)}\}$
		\State $p_R^* \leftarrow \arg\max_{p \in \mathcal{P}_{\text{best}}} \text{Recall@}k(p, \mathcal{M}_{\text{exec}}, \mathcal{D}_{val})$ \Comment{Final Selection on Val}
		
		\Statex \textbf{// Stage II: Detector Prompt Optimization}
		\For{$m = 1$ to $M$ \textbf{in parallel}}
		\State Initialize $\mathcal{P}_m$; Construct $\mathcal{D}_{tr}^{(m)}$, $\mathcal{D}_{val}^{(m)}$
		\For{$t = 1$ to $T$}
		\State $\mathcal{S} \leftarrow \{ \text{F1}(p, m, \mathcal{M}_{\text{exec}}, \mathcal{D}_{tr}^{(m)}) \mid p \in \mathcal{P}_m \}$
		\State $\mathcal{P}_m \leftarrow \textsc{Evolve}(\mathcal{P}_m, \mathcal{S}, \mathcal{M}_{\text{evo}})$
		\EndFor
		\State Select $p_m^*$ using $\mathcal{D}_{val}^{(m)}$ from evolved candidates
		\EndFor
		
		\State \Return $p_R^*$, $\{p_m^*\}_{m=1}^{M}$
	\end{algorithmic}
\end{algorithm}

\begin{algorithm}[t]
	\caption{\textsc{Evolve}: LLM-Driven Prompt Evolution}
	\label{alg:evolve}
	\begin{algorithmic}[1]
		\Require Population $\mathcal{P}$, Fitness scores $\{\mathcal{F}(p)\}_{p \in \mathcal{P}}$, Evolution model $\mathcal{M}_{\text{evo}}$, Elite ratio $\alpha$
		\Ensure Updated prompts $\mathcal{P}'$
		
		\State $\mathcal{P}' \leftarrow$ top-$\lfloor \alpha |\mathcal{P}| \rfloor$ prompts ranked by $\mathcal{F}$ 
		\While{$|\mathcal{P}'| < |\mathcal{P}|$}
		\State Sample $p$ via rank-based selection to maintain diversity
		\State $p' \leftarrow \mathcal{M}_{\text{evo}}(\texttt{mutate}, p, \mathcal{F}(p))$ 
		\State $\mathcal{P}' \leftarrow \mathcal{P}' \cup \{p'\}$
		\EndWhile
		\State \Return $\mathcal{P}'$
	\end{algorithmic}
\end{algorithm}

\subsection{Online Multi-Agent Detection} \label{sec:online} 
Following the two-level CWE hierarchy defined in Section~\ref{sec:cwe}, MulVul employs a coarse-to-fine detection strategy where autonomous agents are equipped with specialized analysis and retrieval tools to ground their decision-making. Figure~\ref{fig:framework} illustrates this tool-augmented architecture.

Given the optimized prompts $p_R^*$ and $\{p_m^*\}_{m=1}^{M}$ from offline preparation, MulVul performs retrieval-augmented multi-agent detection at inference time. Algorithm~\ref{alg:inference} summarizes the procedure.

\subsubsection{Router agent: Global Planning}
Given an input code snippet $x$, the Router agent acts as a dispatcher to predict coarse-grained categories. To overcome the limitations of raw text processing, the agent first employs a Structure Analysis Tool (SCALE) to extract semantic features:
\begin{equation}
	T(x) = \textsc{Tool}_{\text{SCALE}}(x)
\end{equation}
With this structured representation, the agent actively invokes a Global Retrieval Tool to query the knowledge base $\mathcal{K}$ for $r$ cross-category examples:
\begin{equation}
	E_R = \textsc{Tool}_{\text{Global}}(T(x), \mathcal{K}, r).
\end{equation}
The Router agent utilizes these top-$r$ retrieved examples to understand the broad semantic context. The Router agent then takes the optimized prompt $p_R^*$, the original code $x$, and evidence $E_R$ as input, and outputs a ranked list of top-$k$ category predictions:
\begin{equation}
	\mathcal{C}_{\text{top-}k} = \textsc{Router}(p_R^*, x, E_R)
\end{equation}

\subsubsection{Detector agents: Fine-grained Identification}
For each predicted category $c_m \in \mathcal{C}_{\text{top-}k}$, the corresponding Detector agent performs fine-grained vulnerability type identification. To prevent confirmation bias, the Detector agent is equipped with a Contrastive Retrieval Tool. This tool dynamically sources evidence from three distinct pools: in-category positives $\mathcal{K}_m$, clean examples $\mathcal{K}_0$, and out-of-category hard negatives $\mathcal{K}_{\neg m}$. The agent allocates its retrieval budget as $r_{\text{pos}} = r_{\text{neg}} = \lfloor r/3 \rfloor$ and $r_{\text{hard}} = r - r_{\text{pos}} - r_{\text{neg}}$.

Based on this allocation, the agent invokes the tool to construct the context: 
\begin{equation} 
	E_m = \textsc{Tool}_{\text{Contrast}}(T(x), c_m, \mathcal{K}, r) 
\end{equation} 
Each Detector agent then analyzes this contrastive context to produce a prediction:
\begin{equation}
	(\hat{\mathcal{Y}}_m, \hat{\mathcal{E}}_m) = \textsc{Detector}_m(p_m^*, x, E_m)
\end{equation}
where $\hat{\mathcal{Y}}_m$ represents the identified vulnerability types and $\hat{\mathcal{E}}_m$ contains the explanations.  By operating with isolated tools, the agents avoid error cascading. After all invoked Detector agents return their predictions, MulVul aggregates them to produce the final output. 

\begin{algorithm}[t]
	\caption{MulVul Online Detection}
	\label{alg:inference}
	\begin{algorithmic}[1]
		\Require Code snippet $x$, Knowledge base $\mathcal{K}$, Category subsets $\{\mathcal{K}_m\}_{m=1}^{M}$, Router prompt $p_R^*$, Detector prompts $\{p_m^*\}_{m=1}^{M}$ 
		\Ensure Prediction $\hat{\mathcal{Y}}$, Evidence $\hat{\mathcal{E}}$
		
		\Statex \textbf{// Phase I: Coarse-grained Routing}
		\State $T(x) \leftarrow \textsc{Tool}_{\text{SCALE}}(x)$  \Comment{Structure Analysis}
		\State $E_R \leftarrow \textsc{Tool}_{\text{Global}}(T(x), \mathcal{K}, r)$  
		\State $\mathcal{C}_{\text{top-}k} \leftarrow \textsc{Router}(p_R^*, x, E_R)$ 
		
		\Statex
		\Statex \textbf{// Phase II: Fine-grained Detection}
		\State $\hat{\mathcal{Y}} \leftarrow \varnothing$; $\hat{\mathcal{E}} \leftarrow \varnothing$
		\For{$c_m \in \mathcal{C}_{\text{top-}k}$ \textbf{in parallel}}
		\State \textbf{// Detector invokes contrastive tool}
		\State $E_m \leftarrow \textsc{Tool}_{\text{Contrast}}(T(x), m, \mathcal{K}, r)$
		\State $(\hat{\mathcal{Y}}_m, \hat{\mathcal{E}}_m) \leftarrow \textsc{Detector}_m(p_m^*, x, E_m)$ 
		\State $\hat{\mathcal{Y}} \leftarrow \hat{\mathcal{Y}} \cup \hat{\mathcal{Y}}_m$
		\State $\hat{\mathcal{E}} \leftarrow \hat{\mathcal{E}} \cup \hat{\mathcal{E}}_m$
		\EndFor
		
		\Statex \textbf{// Phase III: Aggregation}
		\If{$\hat{\mathcal{Y}} = \varnothing$} \State $\hat{\mathcal{Y}} \leftarrow \{y_0\}$ \EndIf
		\State \Return $\hat{\mathcal{Y}}$ , $\hat{\mathcal{E}} $
	\end{algorithmic}
\end{algorithm}

\section{Evaluation}\label{sec:experiment}
We evaluate MulVul through comprehensive experiments designed to answer the following questions:
\begin{itemize}
	\item \textbf{Q1:} How does MulVul compare with existing LLM-based vulnerability detection methods?
	\item \textbf{Q2:} How does the routing parameter $k$ affect the precision-recall trade-off?
	\item \textbf{Q3:} How do different components contribute to MulVul's performance?
	\item \textbf{Q4:} How does MulVul perform on few-shot CWE types?
\end{itemize}

\subsection{Experimental Setup}\label{sec:setup} 
\paragraph{Dataset.} 
We evaluate on PrimeVul~\cite{ding2024vulnerability}, containing 6,968 vulnerable and 229,764 benign C/C++ functions across 10 categories and 130 CWE types.

\paragraph{Implementation.}
We use GPT-4o as the execution model $\mathcal{M}_{\text{exec}}$ for both Router and Detector agents, and Claude Opus 4.5 as the evolution model $\mathcal{M}_{\text{evo}}$. We use UniXcoder~\cite{guo2022unixcoder} for embedding and FAISS for retrieval.  
\paragraph{Metrics.}
Following~\citet{ding2024vulnerability}, we report Macro-Precision, Macro-Recall, and Macro-F1. Macro-averaging computes metrics independently for each CWE type and then averages them, ensuring equal weight for all types and avoiding dominance by high-frequency vulnerabilities under severe class imbalance.
\paragraph{Baselines.}
We compare our approach with four state-of-the-art methods that span prompting, fine-tuning, and GNN paradigms. 
1) GPT-4o: Prompting-based detection without demonstration examples or fine-tuning.
2) LLM$\times$CPG~\cite{lekssays2025llmxcpg}: LoRA fine-tuned Qwen2.5-32B with CPG-guided context.
3) LLMVulExp~\cite{mao2025towards}: LoRA fine-tuned CodeLlama-7B with chain-of-thought explanations. 
4) VISION~\cite{egea2025vision}: Devign GNN with counterfactual augmentation. 
LLM$\times$CPG and VISION are extended from binary to multi-class classification for fair comparison.

\subsection{Comparison of Vulnerability Detection Effectiveness (Q1)}\label{sec:main_results} 
We compare the vulnerability detection effectiveness of MulVul with existing methods at the coarse-grained category level and fine-grained type level. 
\paragraph{Category-Level Detection.}
Table~\ref{tab:category_results} reports category-level results. MulVul achieves the best overall performance with 50.41\% Macro-F1, outperforming the strongest baseline LLMVulExp by 8.91 points. MulVul also achieves the highest Macro-Precision (44.31\%) while maintaining strong Macro-Recall (58.45\%), indicating accurate category identification with fewer false positives. By contrast, LLM$\times$CPG yields the highest recall (62.81\%) but substantially lower precision (27.44\%), suggesting that expanding candidates improves coverage but induces over-prediction.
\begin{table}[h]
	\centering
	\small
	\setlength{\tabcolsep}{1.5pt} 
	\begin{tabular}{lccc}
		\toprule
		\textbf{Method} & \textbf{Macro-Precision} & \textbf{Macro-Recall} & \textbf{Macro-F1} \\
		\midrule
		GPT-4o & 22.20 & 13.13 & 16.50 \\
		LLM$\times$CPG & 27.44 & \textbf{62.81} & 38.20 \\
		LLMVulExp & 37.88 & 45.88 & 41.50 \\
		VISION & 22.23 & 33.72 & 26.80 \\
		MulVul (Ours) & \textbf{44.31} & 58.45 & \textbf{50.41} \\
		\bottomrule
	\end{tabular}
	\caption{Category-level vulnerability detection effectiveness (\%) on PrimeVul. }
	\label{tab:category_results}
\end{table}

\paragraph{Type-Level Detection.}
Table~\ref{tab:type_results} presents type-level results. The task is markedly harder: all baselines exhibit a sharp precision drop, reflecting that fine-grained types require discriminative evidence beyond generic vulnerability semantics. MulVul achieves 34.79\% Macro-F1, surpassing LLM$\times$CPG by 10.21 points. Importantly, MulVul improves Macro-Precision to 27.90\% while keeping competitive Macro-Recall, yielding a stronger precision--recall trade-off that is critical for practical deployment.
\begin{table}[h]
	\centering
	\small
	\setlength{\tabcolsep}{1.5pt} 
	\begin{tabular}{lccc}
		\toprule
		\textbf{Method} & \textbf{Macro-Precision} & \textbf{Macro-Recall} & \textbf{Macro-F1} \\
		\midrule
		GPT-4o & 4.76 & 3.28 & 3.86 \\
		LLM$\times$CPG & 16.80 & 45.80 & 24.58 \\
		LLMVulExp & 19.31 & 27.60 & 22.72 \\
		VISION & 14.91 & 23.12 & 18.12 \\
		MulVul (Ours) & \textbf{27.90} & \textbf{46.19} & \textbf{34.79} \\
		\bottomrule
	\end{tabular}
	\caption{Type-level vulnerability detection effectiveness (\%) on PrimeVul.}
	\label{tab:type_results}
\end{table}

\subsection{Impact of Routing Parameter $k$ (Q2)}\label{sec:topk} 
The routing parameter $k$ controls the number of candidate categories the Router passes to downstream Detectors, directly affecting the precision-recall trade-off.
\begin{table}[h]
	\centering
	\small
	\begin{tabular}{cccc}
		\toprule
		\textbf{Top-}$k$ & \textbf{Macro-Precision} & \textbf{Macro-Recall} & \textbf{Macro-F1}\\
		\midrule
		1 & \textbf{39.58} & 29.10 & 33.51 \\
		2 & 28.80 & 43.20 & 34.56 \\
		3 & 27.90 & 46.19 & \textbf{34.79} \\
		4 & 26.79 & 47.83 & 34.34 \\
		5 & 26.54 & \textbf{48.37} & 34.27 \\
		\bottomrule
	\end{tabular}
	\caption{Effect of routing parameter $k$ on PrimeVul (Type-level, \%).}
	\label{tab:topk}
\end{table}
\paragraph{Analysis.}
We observe three key patterns. 
First, Macro-Recall consistently increases as $k$ grows. This indicates that allowing the Router to activate multiple candidate categories substantially reduces missed detections, as the true class is more likely to be covered by the expanded Top-$k$ set.
Second, Macro-Precision shows a clear downward trend with larger $k$. As more detectors are triggered, incorrect categories are increasingly introduced, leading to more false positives and thus lower precision. This behavior reflects the inherent trade-off between coverage and noise when expanding the routing space.
Third, Macro-F1 reaches its peak at $k{=}3$ and remains relatively stable beyond this range. Although recall continues to improve for larger $k$, the corresponding degradation in precision offsets these gains, resulting in diminishing overall benefits.  

\subsection{Ablation Study (Q3)}\label{sec:ablation}
To understand the contribution of each component in MulVul, we conduct ablation studies by removing key modules. Table~\ref{tab:ablation} presents the results. 

\begin{table}[H]
	\centering
	\small
	\setlength{\tabcolsep}{1.5pt} 
	\begin{tabular}{lccccc}
		\toprule
		\textbf{Variant} & \textbf{RAG} & \textbf{Agents} & \textbf{Evolution} & \textbf{Macro-F1} & \textbf{$\Delta$} \\
		\midrule
		GPT-4o    & \ding{55} & \ding{55} & \ding{55} & 3.86 & -30.70 \\
		w/o Retrieval & \ding{55} & \ding{51} & \ding{51} & 21.80 & -12.76 \\
		w/o Agents & \ding{51} & \ding{55} & \ding{51} & 28.61 & -5.95 \\
		w/o Evolution & \ding{51} & \ding{51} & \ding{55} & 22.80 & -11.76 \\
		\midrule
		MulVul (Full) & \ding{51} & \ding{51} & \ding{51} & \textbf{34.56} & -- \\
		\bottomrule
	\end{tabular}
	\caption{Ablation study on PrimeVul (Type-level). $\Delta$ denotes the Macro-F1 difference from the full model.}
	\label{tab:ablation}
\end{table}

\paragraph{Analysis.} Retrieval augmentation is the most critical component. Removing evidence retrieval causes the largest performance drop, reducing Macro-F1 from 34.56\% to 21.80\%. This confirms that grounding LLM reasoning with retrieved vulnerability examples from the knowledge base $\mathcal{K}$ is essential for distinguishing semantically similar CWE types. Without concrete code evidence, even well-structured prompts and specialized agents struggle to make accurate fine-grained predictions.
Moreover, cross-model prompt evolution provides substantial gains. 
Replacing evolved prompts with manual templates leads to an 11.76\% F1 drop, demonstrating that our cross-model evolution strategy (Section~\ref{sec:prompt_evolution}) effectively optimizes task-specific instructions. 

\subsection{Performance on Few-Shot CWE Types (Q4)}\label{sec:fewshot}
\begin{figure}[t]
	\centering
	\includegraphics[width=0.95\linewidth]{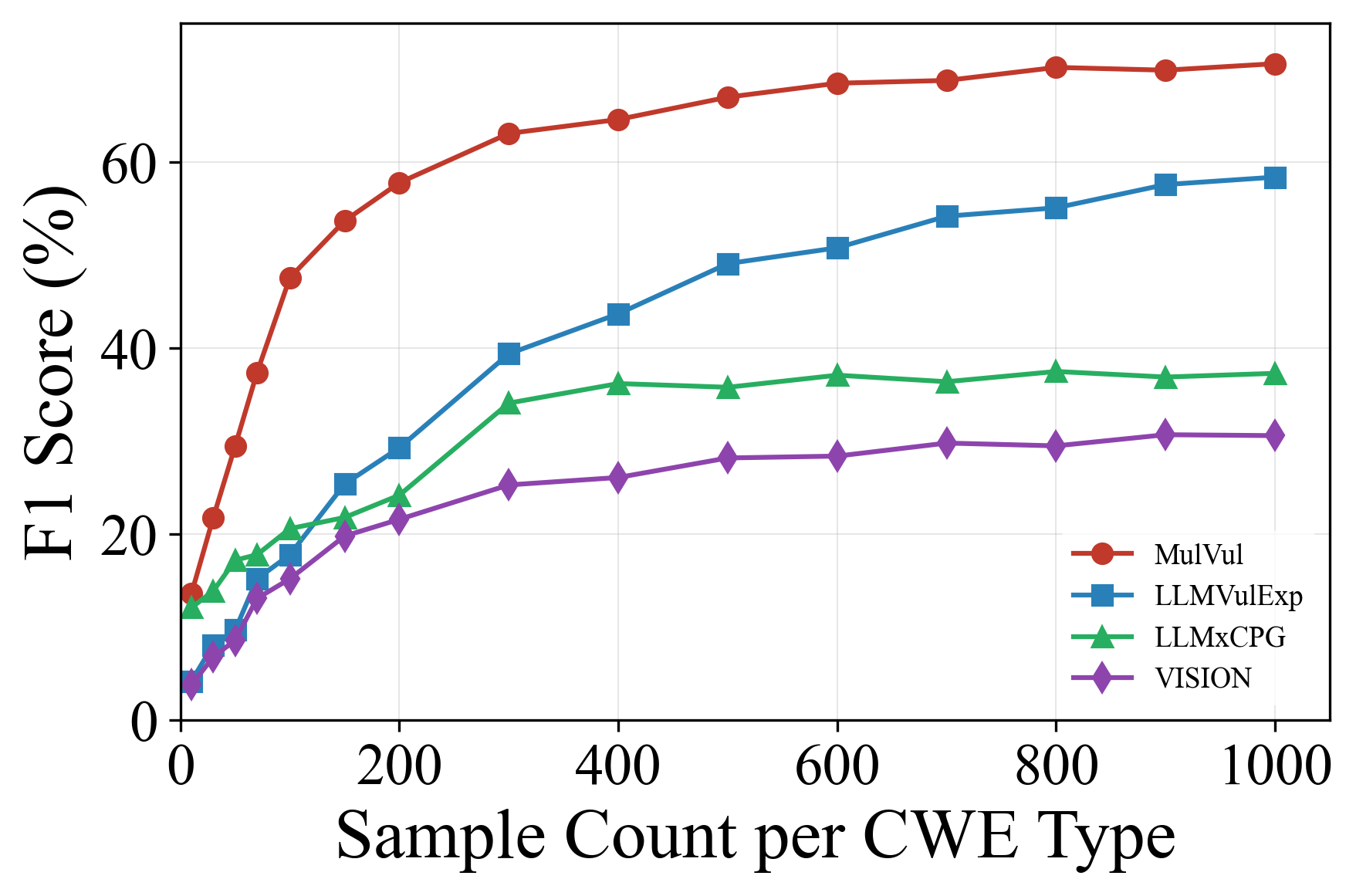}
	\caption{F1 score vs. CWE sample count. MulVul outperforms baselines across all data regimes, with the largest gains on few-shot CWEs.}
	\label{fig:fewshot}
\end{figure}

Real-world vulnerability datasets exhibit severe class imbalance, with many CWE types having only a small number of samples (i.e., few-shot settings). We analyze how methods perform across CWEs grouped by sample count. Figure~\ref{fig:fewshot} visualizes the relationship between CWE sample size and detection performance.

\paragraph{Analysis.}
First, MulVul has a strong few-shot performance. For CWEs with fewer than 100 samples, MulVul achieves approximately 48\% F1, nearly doubling the performance of the best baseline LLMVulExp (25\%). This demonstrates that retrieval augmentation enables effective cross-CWE knowledge transfer. Similar vulnerability patterns from related types provide useful detection signals even when target-type samples are scarce.

Second, MulVul's performance curve rises steeply and plateaus around 300 samples at approximately 63\% F1, while fine-tuning methods (LLM$\times$CPG, VISION) plateau much earlier at lower performance levels (35-38\%). This indicates that MulVul extracts more discriminative information from limited samples, a crucial advantage for practical deployment where many vulnerability types are inherently rare. 

Third, the advantage persists in data-rich regimes. Even for CWE types with over 500 samples, MulVul maintains a 12+ point F1 lead over baselines. This demonstrates that MulVul's coarse-to-fine strategy and architecture are all more beneficial than existing schemes.

See Appendix~\ref{app:cwe_analysis} for per-CWE analysis and quantitative few-shot metrics.

\section{Conclusion}\label{sec:conclusion}
We propose MulVul, a retrieval-augmented multi-agent framework for vulnerability detection. The coarse-to-fine Router-Detector architecture addresses the heterogeneity and scalability challenges in analyzing massive weakness categories. Additionally, cross-model prompt evolution automates the discovery of specialized instructions while mitigating self-correction bias, and SCALE-based contrastive retrieval grounds LLM reasoning. Experiments on PrimeVul demonstrate that MulVul achieves a state-of-the-art 34.79\% Macro-F1 (41.5\% relative improvement), and our evolutionary mechanism yields a 51.6\% performance boost over manual prompt engineering.

\clearpage
\section*{Limitations}
We acknowledge several limitations of our work:
\begin{itemize}
	\item MulVul is evaluated exclusively on PrimeVul containing C/C++ code. Effectiveness on other programming languages (e.g., Java, Python) with different vulnerability patterns and on other benchmarks remains unexplored.
	\item MulVul requires multiple LLM API calls: iterative optimization during offline prompt evolution and $1+k$ calls per sample during online detection. This may limit applicability in resource-constrained or large-scale batch processing scenarios.
	\item Although we claim that cross-model evolution improves generalization, our experiments primarily use GPT-4o as the execution model. The transferability of evolved prompts to other LLMs was not thoroughly evaluated due to scope constraints.
	\item We recognize the potential for misuse associated with automated vulnerability detection tools. While MulVul is designed to aid developers in securing code, malicious actors could theoretically utilize the framework to discover zero-day vulnerabilities in software systems for exploitation. Furthermore, there is a risk of automation bias; developers might develop a false sense of security and reduce manual scrutiny, which is dangerous given that our model inevitably produces false negatives.
\end{itemize}

\bibliography{custom}

\clearpage

\appendix
\section{Few-Shot CWE Performance Analysis}\label{app:cwe_analysis}

This appendix provides a detailed analysis of detection performance on individual CWE types, complementing the aggregated results in Section~\ref{sec:fewshot}. We examine how class imbalance affects each method and quantify MulVul's advantages on few-shot CWE types, i.e., those with limited training samples.

\subsection{Class Imbalance in PrimeVul}
Table~\ref{tab:cwe_distribution} characterizes the dataset's long-tail distribution: 69.6\% of samples concentrate in only 12 CWE types, while 48 types (37\% of all CWEs) collectively contain less than 1\% of samples. This severe imbalance creates few-shot scenarios for many CWE types, where models must generalize from extremely limited examples.

\begin{table}[h]
	\centering
	\small
	\begin{tabular}{lrrc}
		\toprule
		\textbf{Tier} & \textbf{\#CWEs} & \textbf{Samples} & \textbf{Share} \\
		\midrule
		Head ($\geq$5k) & 12 & 140,080 & 69.6\% \\
		Medium (1k--5k) & 16 & 42,554 & 21.2\% \\
		Low (100--1k) & 54 & 16,735 & 8.3\% \\
		Rare (50--100) & 16 & 1,109 & 0.6\% \\
		Very Rare (10--50) & 25 & 656 & 0.3\% \\
		Ext. Rare ($<$10) & 7 & 32 & $<$0.1\% \\
		\midrule
		\textbf{Total} & \textbf{130} & \textbf{201,166} & \textbf{100\%} \\
		\bottomrule
	\end{tabular}
	\caption{CWE distribution in PrimeVul. The bottom four tiers (48 CWE types) represent few-shot scenarios with $<$100 samples each.}
	\label{tab:cwe_distribution}
\end{table}

\subsection{Per-CWE Performance Visualization}
\begin{figure}[h]
	\centering
	\includegraphics[width=\linewidth]{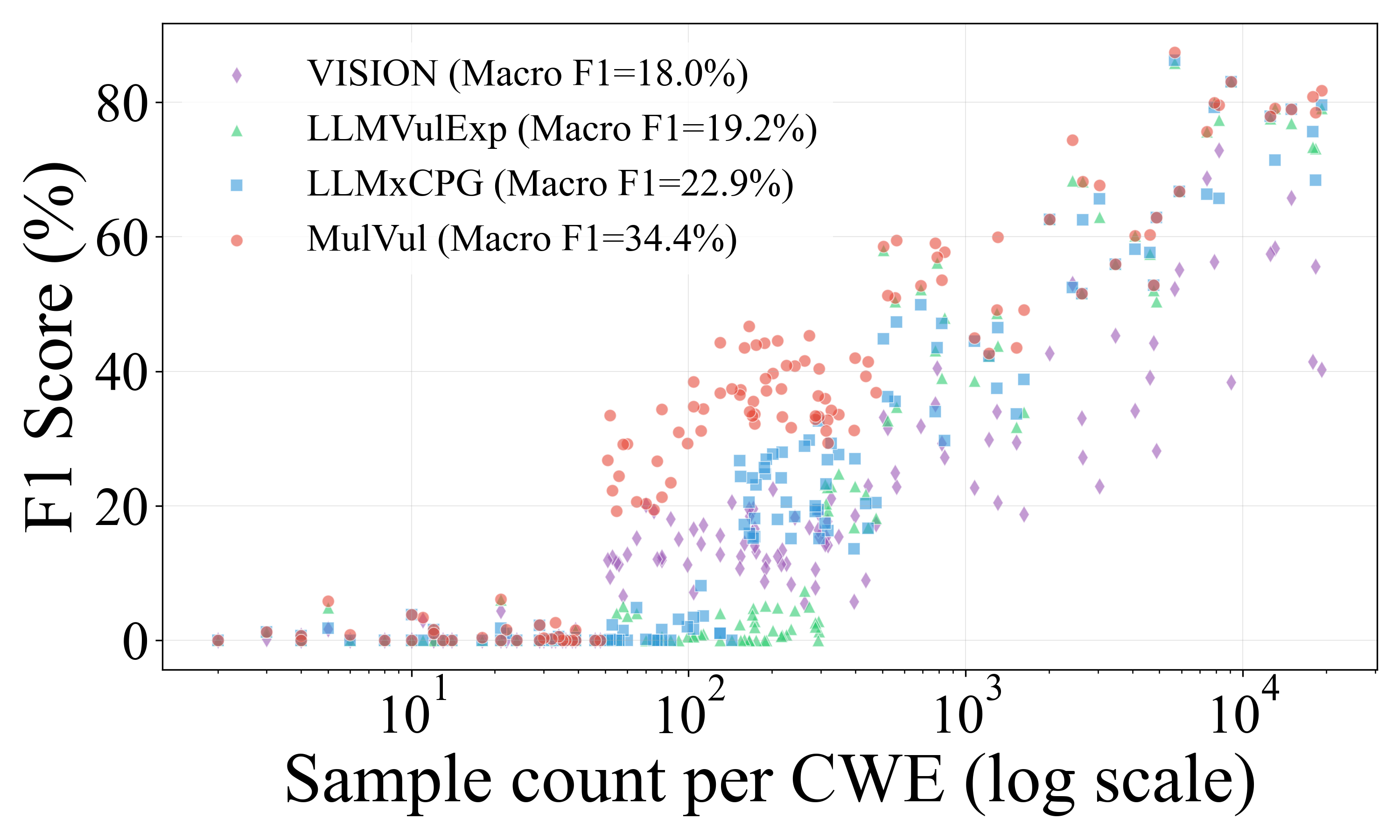}
	\caption{Per-CWE F1 vs. sample count (log scale). MulVul (red) consistently outperforms baselines, especially in the few-shot region (left) .}
	\label{fig:cwe_scatter}
\end{figure}

Figure~\ref{fig:cwe_scatter} plots each method's F1 on every CWE type against its sample count. This visualization reveals how performance scales with data availability and identifies which methods handle few-shot CWEs effectively.

\subsection{Few-Shot Performance Metrics}
To quantify the few-shot detection capability, we define four metrics focusing on CWE types with $<$500 samples. Table~\ref{tab:fewshot_metrics} presents the results.
\begin{table}[h]
	\centering
	\small
	\setlength{\tabcolsep}{4pt}
	\begin{tabular}{lcccc}
		\toprule
		\textbf{Metric} & \textbf{MulVul} & \textbf{LLM$\times$CPG} & \textbf{VulExp} & \textbf{VISION} \\
		\midrule
		Tail F1 $\uparrow$ & \textbf{0.228} & 0.095 & 0.036 & 0.094 \\
		Fail Thresh. $\downarrow$ & \textbf{51} & 153 & 311 & \textbf{51} \\
		Tail Cov. $\uparrow$ & \textbf{65.6\%} & 40.9\% & 11.8\% & 55.9\% \\
		Gini Coef. $\downarrow$ & \textbf{0.396} & 0.580 & 0.695 & 0.495 \\
		\bottomrule
	\end{tabular}
	\caption{Few-shot performance metrics ($\uparrow$: higher is better; $\downarrow$: lower is better). VulExp = LLMVulExp.}
	\label{tab:fewshot_metrics}
\end{table}

\noindent\textbf{Metric Definitions:}
\begin{itemize}[leftmargin=*, itemsep=1pt, topsep=2pt]
	\item \textit{Few-Shot F1}: Average F1 on CWEs with $<$500 samples.
	\item \textit{Min. Samples}: Minimum samples needed for F1 $>$ 0 (data efficiency).
	\item \textit{Coverage}: Fraction of few-shot CWEs achieving F1 $>$ 0.1 (detection breadth).
	\item \textit{Gini Coefficient}: F1 distribution inequality across CWEs (0 = uniform, 1 = skewed).
\end{itemize}

\paragraph{Analysis}
First, MulVul achieves the highest few-shot F1.
MulVul achieves 0.228 F1 on few-shot CWEs, which is 2.4$\times$ higher than LLM$\times$CPG (0.095) and 6.3$\times$ higher than LLMVulExp (0.036). This confirms that retrieval augmentation enables cross-CWE knowledge transfer: when a CWE type has few samples, MulVul leverages similar patterns from the knowledge base. In contrast, fine-tuning methods need substantial data to learn discriminative features, while retrieval-based methods generalize from analogous examples.

Moreover, MulVul shows the most balanced performance. The Gini coefficient measures how uniformly F1 scores are distributed across CWE types. MulVul's lowest Gini (0.396) indicates consistent performance regardless of class frequency, while LLMVulExp's high Gini (0.695) reveals heavy bias toward frequent classes. This balance is essential for Macro-F1 optimization under class imbalance.

\section{Case Study: Impact of Prompt Evolution}
To analyze how MulVul improves prompt robustness, Figure~\ref{fig:prompt_comparison} visually contrasts the manually designed prompt (Stage 0) with the final evolved prompt (Stage T).
\begin{figure*}[t]
	\centering
	\footnotesize
	\begin{minipage}{0.48\textwidth}
		\begin{tcolorbox}[
			colback=initgray,
			colframe=gray!60!black,
			title=\textbf{Initial Prompt (Manual Design)},
			boxrule=0.8pt,
			arc=2pt
			]
			\textbf{Role:} You are a security expert specializing in detecting coding vulnerabilities. \\
			
			\textbf{Instructions:}\\
			1. Identify which patterns from the evidence examples appear in the target code.\\
			2. If the code shares vulnerable patterns with the examples, classify accordingly.\\
			3. Only mark as Benign if NO similar patterns exist. \\
			
			\textbf{Categories:} Memory, Injection, Logic, Input, Crypto, Benign \\
			
			\textbf{Target Code:} \{code\} \\
			\textbf{Evidence:} \{evidence\} \\
			
			\textbf{Output (JSON):}
			\{ "predictions": [ \{ "category": "...", "confidence": 0.85, "reason": "..." \} ] \}
		\end{tcolorbox}
		\vspace{0.1cm}
		\centering \small \textit{(a) Baseline prompt lacks definitions and explicit constraints.}
	\end{minipage}
	\hfill
	\begin{minipage}{0.48\textwidth}
		\begin{tcolorbox}[
			colback=evoblue,
			colframe=blue!60!black,
			title=\textbf{Evolved Prompt (Optimized by MulVul)},
			boxrule=0.8pt,
			arc=2pt
			]
			\textbf{Role:} You are a \textbf{\textcolor{evodark}{senior security analyst}}. Determine vulnerability by \textbf{\textcolor{evodark}{explicitly comparing against confirmed patterns}}. \\
			
			\textbf{Constraints:}
			- \textbf{\textcolor{evodark}{Do NOT infer vulnerabilities beyond these patterns.}}\\
			- \textbf{\textcolor{evodark}{Do NOT speculate about hypothetical vulnerabilities.}} \\
			
			\textbf{Category Definitions (Key Signals):}
			- \textit{Memory}: Direct memory manipulation \textbf{\textcolor{evodark}{without bounds}}.\\
			- \textit{Injection}: User data reaches execution context.\\
			- \textit{Input}: \textbf{\textcolor{evodark}{Distinction: Affects data integrity but does NOT execute code.}}\\
			- \textit{Logic}: Code "works as written" but violates security assumptions. \\
			
			\textbf{Error Prevention Hints:}
			- \textbf{\textcolor{evodark}{Injection vs Input:}} Injection executes instructions; Input flaws only mishandle data.\\
			- \textbf{\textcolor{evodark}{Benign vs Vulnerable:}} If no strong pattern match, default to Benign. \\
			
			\textbf{Output Format (STRICT JSON):} ...
		\end{tcolorbox}
		\vspace{0.1cm}
		\centering \small \textit{(b) Evolved prompt incorporates negative constraints, disambiguation rules, and specific signals.}
	\end{minipage}
	
	\caption{Comparison of the Router agent's prompt before and after Cross-Model Evolution. The \textbf{Initial Prompt} (a) relies on generic instructions, while the \textbf{Evolved Prompt} (b) introduces semantic disambiguation (e.g., Injection vs. Input) and negative constraints (e.g., ``Do NOT speculate") to mitigate hallucinations. High-impact additions are highlighted in \textbf{\textcolor{evodark}{bold blue}}.}
	\label{fig:prompt_comparison}
\end{figure*}

First, MulVul enables a shift from implicit to explicit definitions. As shown in Figure~\ref{fig:prompt_comparison}(a), the initial prompt lists categories without definitions, relying entirely on the LLM's parametric knowledge. This often leads to confusion between conceptually similar types, such as \textit{Injection} (CWE-74) and \textit{Input Validation} (CWE-20). In contrast, the evolved prompt in Figure~\ref{fig:prompt_comparison}(b) explicitly injects discriminative boundaries (e.g., \textit{``Input flaws affect data integrity but do NOT execute code"}). This change, driven by the error feedback loop during evolution, significantly improves the Router's classification precision.

Second, MulVul mitigates false positives through negative constraints. A major challenge in vulnerability detection is the high false positive rate caused by LLMs' ``hallucinating" flaws in benign code. The evolutionary process introduced negative constraints, highlighted in bold blue in Figure~\ref{fig:prompt_comparison}(b) (e.g., \textit{``Do NOT infer vulnerabilities beyond these patterns"}). These ``stop words" act as guardrails, forcing the agent to output \textit{Benign} when evidence is insufficient, thereby reducing the False Positive Rate.

Third, MulVul adds an error prevention mechanism. The evolved prompt includes a novel \textit{``Error Prevention Hints"} section. This suggests that the Executor LLM (GPT-4o) successfully identified recurring confusion patterns in early iterations and the Generator LLM (Claude) synthesized these observations into explicit ``Chain-of-Thought" rules (e.g., \textit{Memory vs. Logic}) to guide future reasoning.

\section*{Acknowledgments}
We used large language models (Gemini, Claude, and GPT-5.2) to assist with grammar checking, polishing, and improving the clarity of the writing. All technical contributions, experimental design, implementation, and analysis were conducted entirely by the authors. The authors take full responsibility for the content of this paper.
\end{document}